\documentclass[12pt]{elsart}
\usepackage{graphicx}
\usepackage{amsmath,amssymb,amsfonts,epsf,epsfig}

\newcommand{\bi}{\begin{itemize}}
\newcommand{\ei}{\end{itemize}}
\newcommand{\be}{\begin{eqnarray}}
\newcommand{\ee}{\end{eqnarray}}
\newcommand{\ba}{\begin{array}}
\newcommand{\ea}{\end{array}}
\newcommand{\bc}{\begin{center}}
\newcommand{\ec}{\end{center}}
\newcommand{\bt}{\begin{tabular}}
\newcommand{\btab}{\begin{table}}
\newcommand{\et}{\end{tabular}}
 
\begin{document}
\begin{frontmatter}
\title{Black Holes and Generalized Scalar Field}
\author[mj]{Marcin Jankiewicz},
\author[as]{Anjan A. Sen}

\address{Department of Physics and Astronomy,\\ Vanderbilt University, Nashville, TN ~~37235}
\thanks[mj]{m.jankiewicz@vanderbilt.edu}
\thanks[as]{anjan.sen@vanderbilt.edu}

\begin{abstract}
We study the possibility of occurrence of scalar hair with a non-canonical kinetic term for a static, spherically symmetric asymptotically flat black hole spacetime. We first obtain a general equation for this purpose and then consider various examples for the kinetic term $F(X)$ with $X = -{1\over{2}}\partial^{\mu}\phi\partial_{\mu}\phi$. Our study shows that for a tachyon field with a positive potential, which naturally arises in open string theory, asymptotically flat a static black hole solution does not exist.  

\end{abstract}
\end{frontmatter}

\section{Introduction}
The existence or nonexistence of the scalar hair has been the topic of active research for more than three decades (see \cite{heusler1} for a nice review). It actually started with the famous statement by J.A. Wheeler ``a black hole has no hair'' \cite{wheeler}. This is due to the nature of stationary black holes as they are completely characterized by three conserved quantities, mass, electric charge and angular momentum, which can be measured at asymptotic infinity. This is the so called ``no hair theorem'' for black holes.  It suggests that all matter fields present in the black hole spacetime should be either radiated to infinity or vanish inside of the black hole, except the three conserved charges mentioned above. This result has been proven for the vacuum case \cite{wald}, for Einstein-Maxwell theory \cite{wald} as well as for several minimally \cite{minimal} and non-minimally coupled scalar field theories \cite{nonminimal}.

In recent years, there have been a number of investigations studying the black hole solutions with non-linear matter fields. The discovery of black hole solutions in Einstein-Yang-Mills theory \cite{bizon}, Einstein-Skyrme  theory \cite{skyrme} , Einstein-Yang-Mills-Higgs theory \cite{higgs} as well as Einstein-non-Abelian-Proca theory \cite{proca}, are all examples of violation of the ``no hair theorem''. Also the Bronikov-Melnikov-Bocharova-Bekenstein solution \cite{Bekenstein:1974sf}, which corresponds to a spherically-symmetric black hole solution with scalar field conformally coupled to gravity, is another example of scalar hair, although it was later shown that in this configuration, scalar field diverges at the horizon. This and the fact that the energy-momentum tensor of this field is ill-defined at the horizon, is enough to dismiss this as a candidate for a regular black hole solution.

The obstacle of obtaining a regular black hole solution with a self interacting scalar field, can be overcome by introducing a cosmological constant $\Lambda$  and a conformal coupling \cite{conformal}. For a minimally coupled scalar field, one needs to have $\Lambda < 0$ \cite{ads}, which allows existence of a black hole with scalar hair and nontrivial topology on the horizon \cite{zanelli}.

Recently, an effective scalar field theory governed by a Lagrangian with a non-canonical kinetic term 
($\mathcal{L}=-V(\phi)F(X)$, where $X = -{1\over{2}}\partial^{\mu}\phi\partial_{\mu}\phi$), has attracted considerable attention. One example of such a field is the tachyon in open string field theory \cite{Sen:2002nu,Sen:2002in}. Such a model can lead to a late time accelerated expansion and is called  ``k-essence'' \cite{kessence,Padmanabhan:2002sh,Bagla:2002yn}. It is worth noting that models such as the Generalized Chaplygin Gas (GCG) \cite{gcg} which has been proposed to unify the dark matter and dark energy of the universe, may be seen as a special case of k-essence. A Lagrangian with a non-canonical kinetic term has also been investigated for an early universe inflationary scenario and is termed ``k-inflation'' \cite{kinflation}.

In this letter, we study the existence of scalar hair for Lagrangians containing non-canonical kinetic terms for a static asymptotically flat spherically-symmetric black hole spacetime with a regular horizon. We have considered various examples for the Lagrangian and our study shows that with a suitably chosen Lagrangian, it is possible to have a non-trivial asymptotically flat static black hole spacetime with scalar hair.

\section{Asymptotically Flat Black Hole Solutions}
We restrict ourselves to the minimally coupled case and start with a general action given by

\begin{equation}
{\mathcal{S}} = \int d^{4}x \sqrt{-g}\left[{1\over{2}}R + F(X,\phi)\right],
\end{equation}
where $R$ is the scalar curvature, $F(X,\phi)$ is a general function of matter field $\phi$ and $X = - {1\over{2}}\partial^{\mu}\phi\partial_{\nu}\phi$. We have set gravitational constant $\kappa^2 = 8\pi G = 1$. We shall use a sign notation $(-,+,+,+)$ for the metric.

Next we consider the most general static, spherically symmetric black hole spacetime with a regular horizon, whose exterior is given by the metric

\begin{equation}\label{st}
ds^{2} = - e^{-2\delta(r)} A(r) dt^{2} + A^{-1}(r) dr^{2} + r^{2}(d\theta^{2} + \sin^{2}{\theta}d\Phi^{2})\,, 
\end{equation}
where $A(r)$ and $\delta(r)$ are some arbitrary functions of the radial coordinate $r$ only. One can think of $\delta(r)$ as an additional redshift beyond the usual one resulting from the geometry of the static hypersurface.\\
Typically one can parametrize $A(r)$ as

\begin{equation}
A(r)= 1 - {2m(r)\over{r}}\,.
\end{equation}
The existence of a regular horizon at $r= r_{H}$ demands that $2m(r_{H}) = r_{H}$ and $\delta(r_{H})$ is finite. The condition of asymptotic flatness requires that $\mu \rightarrow 1$ and $\delta(r) \rightarrow 0$ as $r \rightarrow \infty$.

The matter field $\phi$ should also respect the symmetries of the spacetime and hence is only a function of radial coordinate $r$. One can now calculate the components of the energy-momentum tensor $T^{\mu}_{\,\,\nu}$ for the scalar field $\phi$

\begin{eqnarray}
T^{t}_{\,\,t} &=& T^{\theta}_{\,\,\theta} = T^{\Phi}_{\,\,\Phi} = F(X,\phi)\,,\nonumber\\
T^{r}_{\,\,r} &=& -2 X F_{X} + F(X,\phi)\,,
\end{eqnarray}
where the subscript 'X' represents derivative with respect to $X$. Einstein's equations $G^{\mu}_{\,\,\nu} = T^{\mu}_{\,\,\nu}$ in this case become

\begin{eqnarray}
{A^{'}}(r) &=& r T^{t}_{\,\,t} + {{1-A(r)}\over{r}}\,,\nonumber\\
{\delta^{'}(r)} &=& {r\over{2A(r)}} ( T^{t}_{\,\,t} - T^{r}_{\,\,r})\,,
\end{eqnarray}
where the prime represents derivative with respect to $r$. The equation of motion is 

\begin{equation}
  \left[e^{-\delta(r)}T^{r}_{\,\,r}\right]^{'} = -{e^{-\delta(r)}\over{2A(r)r}}\left[(T^{t}_{\,\,t} -T^{r}_{\,\,r}) + A(r)(2T - 3T^{t}_{\,\,t} - 5T^{r}_{\,\,r})\right]\,,
\end{equation} 
which can be obtained from the Bianchi identity $T^{\mu}_{\,\,\nu;\mu} = 0$. 
This equation closes the system of equations for the scalar field. Here $T$ stands for the trace of the energy momentum tensor for the scalar field.

One can use the expression for the $T^{\mu}_{\,\,\nu}$'s from above to write this equation as

\begin{equation}\label{master}
\left[e^{-\delta(r)}T^{r}_{\,\,r}\right]^{'} = -{{2 e^{-\delta(r)} \phi^{'2}}\over{r}}\left[1-{3m(r)\over{2r}}\right]F_{X}(X,\phi)\,.
\end{equation}
 This is a key result. This equation generalizes the previous one by Sudarsky \cite{Sudarsky:1995zg} for a minimally couple scalar field with canonical kinetic energy term to one for the non-canonical kinetic term.  One can see that the term inside the square bracket on a r.h.s of the above equation is always positive outside the horizon, i.e., at $r_{H} = 2m(r_{H})$. Hence the action $F(X,\phi)$ determines whether the term $e^{-\delta(r)}T^{r}_{\,\,r}$ is an increasing or decreasing function outside the horizon. Asymptotic flatness requires that it should vanish as $r \rightarrow \infty$. 

Also at $r=r_{H}$, $X = -{1\over{2}}g^{rr}\phi^{'2} = 0$. We assume that, both $F$ and $F_{X}$ are regular at $r=r_{H}$, property which is essential for existence of a regular horizon. Further, we set $T^{t}_{t} = T^{r}_{r} = F = -\rho$ at $r=r_{H}$, where $\rho$ is the energy density for the scalar field. 

Hence to satisfy the Weak Energy Condition (WEC) for the scalar field $\rho > 0$ at $r=r_{H}$, the term $e^{-\delta(r)}T^{r}_{\,\,r}$ on the l.h.s of the above equation is negative at the horizon $r=r_{H}$. Hence it depends solely on the function $F(X,\phi)$ whether one can have asymptotic flatness. With $e^{-\delta(r)}T^{r}_{\,\,r} < 0$ at the  horizon, this condition can be fulfilled for $F_{X} < 0$, as in this case, $e^{-\delta(r)}T^{r}_{\,\,r}$ will be an increasing function outside horizon and can reach zero asymptotically. On the other hand, for $F_{X} > 0$, function $e^{-\delta(r)}T^{r}_{\,\,r}$ will be more negative outside horizon and can not vanish asymptotically.

We shall now discuss different examples for $F(X,\phi)$ to see whether a non-trivial scalar hair exists outside of the regular horizon of an asymptotically flat black hole spacetime.

\subsection{$F(X,\phi) = X - V(\phi)$ with positive $V(\phi)$.}
 This is the standard scalar field case with a canonical kinetic term, that has been widely discussed in the context of scalar hair for black holes. In this case $F_{X}(X,\phi) = 1$, and equation (\ref{master}) becomes

\begin{equation}
\left[e^{-\delta(r)}T^{r}_{\,\,r}\right]^{'} = -{{2 e^{-\delta(r)} \phi^{'2}}\over{r}}\left[1-{3m(r)\over{2r}}\right]\,.
\end{equation}
As r.h.s of this equation is always negative outside the horizon $r_{H} = 2 m(r_{H})$, $e^{-\delta(r)}T^{r}_{\,\,r}$ can not vanish asymptotically and hence one can not have non-trivial scalar field outside of the horizon. This shows that for a asymptotically flat black hole with a regular horizon, one cannot have a non-trivial canonical scalar field with a non-zero potential in the spacetime outside the horizon. This result, has been obtained earlier by Sudarsky \cite{Sudarsky:1995zg}.\\
In our subsequent examples, we shall show that for certain scalar field Lagrangian, it is indeed possible to have the scalar hair in a region outside of the horizon of the asymptotically flat black hole.

\subsection{$F(X,\phi) = -X - V(\phi)$ with positive $V(\phi)$.} 
This is an example of a phantom field \cite{Caldwell:1999ew}. In cosmology, this kind of a field has attracted a lot of attention in recent years, since it has been shown that it can serve as a possible explanation of the present accelerating phase of the universal expansion. The observational data predicts that the equation of state of the dark energy, i.e., $p=w\rho$, responsible for a description of the accelerating universe, has $w$ less than $-1$ and phantom field with the above Lagrangian can be one such candidate for this component. 

In this case as $F_{X}(X,\phi) = -1$, the r.h.s of equation (\ref{master}) is positive and one can have the asymptotic flatness condition satisfied as $e^{-\delta(r)}T^{r}_{\,\,r}$ can vanish asymptotically.    

\subsection{$F(X,\phi) = f(X) - V(\phi)$ with positive $V(\phi)$.} u
This kind of Lagrangian for the scalar field has been recently considered by Mukhanov and Vikman \cite{Mukhanov:2005bu} to show that a non-trivial kinetic term results in substantial contributions of the gravitational waves to the Cosmic Microwave Background (CMB) fluctuations, leading to a larger B-mode CMB polarization, thereby making the prospects of detection of gravity wave in future experiments much more promising. Let us consider the case $f(X) = X^{\alpha}$. As $X < 0$ outside the horizon, $F_{X} < 0$ for $\alpha$ odd positive integer and for $\alpha$ even negative integer. Therefore in these two cases, it is possible to have asymptotically flat black hole solutions with a non-trivial scalar field outside the horizon. For $\alpha$ even positive integer and also for odd negative integer, $F_{X} >0$, and spacetime (\ref{st}) is not asymptotically flat.

\subsection{$F(X,\phi) = -V \sqrt{1-X}$ with positive $V(\phi)$.}
This is the Lagrangian for a Born-Infeld tachyon field which has attracted lot of attention in cosmology recent times. It can act as dark energy as well as a potential candidate for a unified candidate for dark matter and dark energy \cite{Padmanabhan:2002sh,Bagla:2002yn}. This field has also been extensively studied for its possible role of as an inflaton candidate and in connections with tachyons in open string field theory \cite{Sen:2002nu,Sen:2002in}.
In this case $F_{X} = {V\over{2}}(1-X)^{-1/2} > 0$ and hence $e^{-\delta(r)}T^{r}_{\,\,r}$ is a decreasing function outside the horizon and it is not possible to obtain a asymptotically flat solution.

\subsection{$F(X,\phi) = -V (1-X)^{\alpha}$ with positive $V(\phi)$.} 
This is a generalized version for the tachyon Lagrangian. Where $F_{X}$ can be negative for $\alpha < 0$ resulting in a decreasing  $e^{-\delta}T^{r}_{\,\,r}$ outside the horizon. Hence one can obtain a asymptotically flat black hole spacetime.

\section{Conclusions}
Scalar fields with non-canonical kinetic terms have gained in interest in recent years. This is particularly due to the importance for cosmology. Recent investigation show that this type of scalar fields are very useful in modeling the late time acceleration of the universe, as predicted by the latest SN1a observations.  Scalar fields with non-canonical kinetic terms are also common in string theory and a tachyon with Born-Infeld type action arising in open string theory is one such example. 

Until now, no attempt has been made to study the existence of a scalar hair with a non-canonical kinetic term in a model independent way, for a static, spherically-symmetric black hole spacetime (see \cite{olivera} for such an analysis with a specific non-canonical kinetic energy term). This work is the first general study in this direction. Generalizing the work by Sudarsky for a minimally coupled scalar field with a canonical kinetic term, we derive a general equation for studying the existence of scalar hair for non-canonical kinetic term. Equation (\ref{master}) is our main result. Using this equation, we study different examples for the Lagrangian of a scalar field containing s non-canonical kinetic term. It shows that for a tachyon with positive potential, which naturally shows up in open string theory, it is not possible to obtain a asymptotically flat black hole spacetime which is the relevant boundary condition for the spherically-symmetric spacetime. For other cases, existence of the asymptotically flat solutions depends on the choice of the kinetic function $F(X)$.

\section*{Acknowledgments}
This work is supported by U.S. DoE grant $\#$ DE-FG05-85ER40226. The authors are grateful to Thomas W. Kephart and Narayan Banerjee for useful comments and suggestions.

\end{document}